\documentclass[nofootinbib,aps,a4paper,11pt,letterpaper,superscriptaddress,twocolumn,showkeys,times,eqsecnum]{revtex4-2}
\usepackage{amsmath}
\usepackage{amsfonts}
\usepackage{graphicx}
\usepackage{color}
\usepackage{amsthm}
\usepackage{amsmath}
\usepackage{braket}
\usepackage{amsmath}
\usepackage{dcolumn}
\usepackage{bm,url}
\usepackage{mathrsfs}
\usepackage[linktocpage]{hyperref}
\usepackage{subfigure}
\usepackage{microtype}
\usepackage{amsfonts}
\usepackage[left=2.1cm,right=2.1cm,top=2cm,bottom=2cm]{geometry}
\usepackage[usenames,dvipsnames,svgnames]{xcolor}  
\usepackage{hyperref}   
\definecolor{oxfordblue}{rgb}{0.0, 0.13, 0.28}
\definecolor{burgundy}{rgb}{0.5, 0.0, 0.13}
\definecolor{darkolivegreen}{rgb}{0.33, 0.42, 0.18}
\definecolor{darkblue}{rgb}{0,0,0.5}
\definecolor{richcarmine}{rgb}{0.84, 0.0, 0.25}
\definecolor{darkblue}{rgb}{0,0,0.5}
\definecolor{bluer}{rgb}{0.00,0.50,0.75}{}
\hypersetup{colorlinks=true, citecolor=red, linkcolor=blue,
urlcolor = purple, filecolor=magenta}

\def\sideremark#1{\ifvmode\leavevmode\fi\vadjust{\vbox to0pt{\vss
			\hbox to 0pt{\hskip\hsize\hskip1em
				\vbox{\hsize1.3cm\tiny\raggedright\pretolerance10000
					\noindent #1\hfill}\hss}\vbox to8pt{\vfil}\vss}}}%
\begin{document}

\newcommand\be{\begin{equation}}
\newcommand\ee{\end{equation}}
\newcommand\bea{\begin{eqnarray}}
\newcommand\eea{\end{eqnarray}}
\newcommand\bseq{\begin{subequations}} 
\newcommand\eseq{\end{subequations}}
\newcommand\bcas{\begin{cases}}
\newcommand\ecas{\end{cases}}
\newcommand{\p}{\partial}
\newcommand{\f}{\frac}

\title{Cosmology with Logarithmic Corrected Horizon Entropy According to the Generalized Entropy and Variable-$G$ Correspondence}
\author{Chen-Hao Wu}
\email{chenhao\_wu@nuaa.edu.cn}
\affiliation{Center for the Cross-disciplinary Research of Space Science and Quantum-technologies (CROSS-Q), College of Physics, Nanjing University of Aeronautics and Astronautics, \\29 Jiangjun Road, Nanjing City, Jiangsu Province 211106, China}

\author{Yen Chin Ong}
\email{Corresponding author: ongyenchin@nuaa.edu.cn}
\affiliation{Center for the Cross-disciplinary Research of Space Science and Quantum-technologies (CROSS-Q), College of Physics, Nanjing University of Aeronautics and Astronautics, \\29 Jiangjun Road, Nanjing City, Jiangsu Province 211106, China}

\begin{abstract}

According to the GEVAG (Generalized Entropy Varying-G) framework, any modification to the Bekenstein-Hawking area law would also lead to a varying-$G$ gravity theory in which the effective gravitational constant $G_\text{eff}$ becomes area-dependent. Among a myriad of generalized entropy functions explored in the literature, of special interest is the logarithmic correction of quantum gravity. In this work, we apply GEVAG to investigate the effect of logarithmic correction on very early-time cosmology, including the conditions for inflation. We found that if the coefficient of the logarithmic correction term is negative, $G_\text{eff}$ becomes twice that of the current value; whereas, a positive coefficient leads to a very small value of $G_\text{eff}$, which may ameliorate the ``arrow of time'' problem. In fact, slow-roll inflation becomes more natural in the latter case. We make some comparisons with the constant-$G$ approach and reveal the advantages of the GEVAG approach. For example, it can evade the sudden singularity that could otherwise arise when the coefficient of the logarithmic correction term is negative. We also check the validity of the generalized second law and comment on the range of the various parameters.

\end{abstract}

\maketitle
\section{Introduction}
The existence of spacetime singularities, particularly the Big Bang singularity, remains one of the most profound challenges in modern cosmology, signaling the breakdown of general relativity (GR) at the Planck scale. It is widely expected that a consistent theory of quantum gravity (QG) will resolve these singularities by modifying the dynamics of spacetime in the high-energy regime. Among the various candidates for such a theory, loop quantum gravity (LQG) \cite{Rovelli:2008zza, Ashtekar:2021kfp} and asymptotic safety gravity (ASG) \cite{Niedermaier:2006wt, Reuter:2012id} represent two leading, yet distinct, approaches.

LQG is a non-perturbative quantization of gravity, which predicts a discrete microstructure of spacetime characterized by a minimum area gap, proportional to the Planck area $A_p=G\hbar/c^3$. (Henceforth, we will set $c=\hbar=1$.) In the cosmological sector, this discreteness leads to a quantum repulsion that resolves the Big Bang into a quantum bounce \cite{Bojowald:2001xe, Ashtekar:2006wn}. On the other hand, ASG, based on the functional renormalization group (FRG), suggests that gravity constitutes a fundamental quantum field theory well-defined in the ultraviolet (UV) via a non-Gaussian fixed point (NGFP) \cite{Reuter:1996cp}. In this scenario, spacetime remains a continuum manifold, but physical quantities exhibit anomalous scaling dimensions at high energies, often interpreted as a dynamical reduction of spectral dimension \cite{Ambjorn:2005db} or a fractal-like structure \cite{Lauscher:2005qz}.

Despite their fundamental differences, i.e., discreteness versus continuum renormalization flow, both theories suggest corrections to the Bekenstein-Hawking entropy-area relation $S = A/4G$.
A ubiquitous feature in semi-classical analyses of both frameworks is the emergence of a logarithmic correction term\footnote{Note that the Planck area $A_p=G$ in natural units, so the $G$ in the logarithmic item represents some minimum area $A_0$, typically chosen as the Planck area $A_p$.} as
\begin{equation}
S = \frac{A}{4G} + \tilde{c} \ln \left( \frac{A}{G} \right) + \dots
\label{en}
\end{equation}
However, the sign and magnitude of the dimensionless coefficient 
$\tilde{c}$ remains a subject of debate. Microstate counting in LQG typically favors a negative coefficient, e.g., $\tilde{c}=-1/2$ \cite{Meissner:2004ju}, whereas arguments based on renormalization group considerations like ASG often imply a positive contribution \cite{Becker:2012js}. Moreover, Ref.\cite{Ma:2022gtm} reports that the positivity assumption of entropy correction does not hold universally from an aspect of string theory; its applicability depends on the specific background and parameters. Likewise, in the effective field theory approach, the sign depends on the various particle species that are present in the theory \cite{Calmet:2021lny}.

The question is: how does such a correction to the horizon entropy affect cosmological evolution, especially in the very early Universe for which quantum corrections are expected to be important? How does this influence the onset of inflation? 
In this work, rather than deriving cosmological dynamics from first principles within a specific QG theory, we adopt a thermodynamic perspective. Following Jacobson’s seminal insight \cite{Jacobson:1995ab} that the gravitational field equations can be derived from the Clausius relation $\delta Q=TdS$ applied to local horizons, it has been shown that modifying the entropy-area relation leads to modified cosmological equations \cite{Cai:2008ys, Sheykhi:2010wm}. However, previous attempts to apply logarithmically corrected entropy to cosmology often assumed the original conservation law $\nabla^{a}T_{a b}=0$ while keeping the gravitational constant $G$ fixed. Here, we argue that it might be more proper to employ the ``Generalized Entropy Varying-G" (GEVAG) correspondence framework \cite{Lu:2024ppa, Lu:2025ayb}. GEVAG was also derived based on Jacobson's method, but did so already at the level of the field equations. It was found that \emph{any} modification to the standard area law necessitates an effective, area-dependent gravitational coupling $G_{\text{eff}}$. In order to maintain the consistency of the Bianchi identity and the thermodynamic derivation, this inevitably leads to a modified continuity equation\footnote{See also Ref.\cite{Tian:2014sca}, in which the author suggested that to establish a self-consistent correspondence between modified gravity theories and the thermodynamics of spacetime, it is necessary to make $G$ dynamical and accordingly we should modify the conservation law.} $\nabla^a (G_{\text{eff}} T_{ab}) = 0$, representing an energy exchange between the geometry and the matter sector. From this perspective, one can also interpret GEVAG as GR but with a non-trivial correction to the matter sector, in which the ``conservation law'' is directly coupled to the geometry not just via the covariant derivative, but also the area-dependence. See also the recent work \cite{Figliolia:2026sma} for a more rigorous treatment of the idea of GEVAG. 

In this work, by applying this thermodynamically consistent framework to the logarithmic entropy correction, we investigate the resulting cosmological evolution. We find that the sign of the correction parameter $\tilde{c}$ leads to two fundamentally different scenarios in the very early Universe, by lowering or increasing the strength of the effective gravitational constant. This has ramifications for various physics, including the conditions of slow-roll inflation and the arrow of time problem.

This paper is organized as follows: In Sec.(\ref{II}), we review the GEVAG framework and present the basic equations of cosmological dynamics. Then we discuss the diverse very early-time cosmological behaviors introduced by the coefficient of different signs in Sec.(\ref{III}). Inflation is discussed in Sec.(\ref{inflation}).
After which, in Sec.(\ref{IV}), we check the thermodynamic consistency and discuss the preference of parameter selection. Finally, we end with some discussions in Sec.(\ref{V}).

\section{GEVAG FRAMEWORK AND COSMOLOGICAL DYNAMICS}\label{II}
In this section, we first provide a quick review of GEVAG. The readers should refer to \cite{Lu:2024ppa, Lu:2025ayb} for more detailed discussions. Following Jacobson's seminal argument \cite{Jacobson:1995ab}, GEVAG assumes that the Clausius relation $\delta Q=TdS$ holds for all local causal horizons, as well as black hole and cosmological horizons. Here, 
$\delta Q$ is the energy flux crossing the horizon, $T$ and $S$ are the temperature and entropy associated with the horizon, respectively.

The heat flux $\delta Q$ flowing across the horizon is defined by the energy-momentum tensor  $T_{ab}$ of the matter as
\begin{equation}
\delta Q = - \kappa \int_\mathcal{H} \lambda T_{ab} k^a k^b \text{d}\lambda \text{d}A,
\end{equation}
where $A$ is the area of the horizon $\mathcal{H}$ with $k^a$ denoting the tangent vector of the horizon generators, $\kappa$ is the surface gravity, and $\lambda$ is the affine parameter.

On the geometric side, the changes in the horizon area $\delta A$ are governed by the Raychaudhuri equation. The area variation is determined by the Ricci curvature $R_{ab}$ via the standard formula
\begin{equation}
\delta A = - \int_\mathcal{H} \lambda R_{ab} k^a k^b \text{d}\lambda \text{d}A.
\end{equation}

In GR, the horizon entropy is assumed to follow the Bekenstein-Hawking area law $S=A/4G$. If this is generalized to $S=f(A)/4G$, where $f(A)$ is a differentiable function of the area, then the variation of entropy is 
\begin{equation}
\delta S = \frac{\text{d}S}{\text{d}A} \delta A = \frac{f'(A)}{4G} \delta A,
\label{dS_variation}
\end{equation}
where the prime denotes the derivative with respect to $A$.

Substituting these expressions into the thermodynamic relation $\delta Q=TdS$, and identifying the temperature with essentially the surface gravity $T=\kappa/2 \pi$, we obtain
\begin{equation}
-\kappa \int T_{ab} k^a k^b \text{d}\lambda \text{d}A = - \frac{\kappa}{2\pi} \frac{f'(A)}{4G} \int R_{ab} k^a k^b \text{d}\lambda \text{d}A.
\end{equation}
Since this relation must hold for all null vectors $k^a$, the integrands must be the same. This leads to the condition
\begin{equation}
T_{ab} k^a k^b = \frac{f'(A)}{8\pi G} R_{ab} k^a k^b.
\end{equation}
This implies that $8\pi G T_{ab}$ can differ at most by a term proportional to the metric $g_{ab}$, given that $g_{ab}k^a k^b=0$ for null vectors. The resulting field equations can be written as
\begin{equation}
R_{ab} - \frac{1}{2}R g_{ab} + \Lambda g_{ab} = 8\pi G_{\text{eff}} T_{ab},
\label{modified_Einstein_Eq}
\end{equation}
where the cosmological constant $\Lambda$ appears as an integration constant, just like in the original derivation of Jacobson. Crucially, the gravitational constant $G$ is replaced by an effective gravitational coupling
\begin{equation}
G_{\text{eff}}(A) = \frac{G}{f'(A)}.
\label{Geff_general_definition}
\end{equation}
This relation suggests that the strength of the gravitational interaction is not a fundamental constant but runs with the geometric scale of the horizon, or more specially the energy scale. Unlike the case of GR in which $G_\text{eff}\equiv G$ has no area dependence, a generic modification of $A$ to $f(A)$ leads to essentially a non-local $G_\text{eff}$, since area-dependence means that $G_\text{eff}$ is quasi-local. It is still not clear what this means if say, the spacetime has no horizon, or if the spacetime has multiple horizons. See \cite{Lu:2024ppa} for some speculations. In this work, we only focus on cosmological horizon so there is no ambiguity.

Note that in the GEVAG framework, the left-hand side of the field equation is just like that of GR. Thus Bianchi identity would necessitate that one must modify the conservation law of the energy-momentum tensor to $\nabla^a (G_{\text{eff}} T_{ab}) = 0$ rather than $\nabla^a T_{ab} = 0$ alone. Such a modification to the conservation law is not new, and is quite common for varying-$G$ theories. See, for example, \cite{Nagy:2022vei, Tian:2014sca}.

We now focus on the specific case of logarithmic entropy correction, which as mentioned in the Introduction, is a generic prediction of most QG approaches. The entropy is given by Eq.\eqref{en}, and we only keep the leading correction term in this work, so that
\begin{equation}
S = \frac{A}{4G} + \tilde{c} \ln \left( \frac{A}{G} \right),
\end{equation}
where $\tilde{c}$ is a model-dependent dimensionless parameter. This corresponds to the function $f(A)=A+4G \tilde{c} \ln \left( A/G \right)$. Differentiating with respect to $A$, we find
\begin{equation}
f'(A) = 1 + \frac{4G\tilde{c}}{A}.
\end{equation}
Substituting this into Eq.\eqref{Geff_general_definition}, the effective gravitational constant becomes
\begin{equation}
G_{\text{eff}} = \frac{G}{1 + \frac{4G\tilde{c}}{A}}.
\label{Geff_A}
\end{equation}
For a flat FLRW universe, the apparent horizon radius is the Hubble radius $R_H=1/H$, where $H$ is the Hubble parameter denoted as $H(t)=\dot a(t)/a(t)$ (overdot denotes the derivative with respect to cosmic time). The area of the apparent horizon is therefore $A=4\pi/H^2$. Substituting this geometric relation into Eq.\eqref{Geff_A}, we find that $G_{\text{eff}}$ becomes a function of the Hubble parameter $H$ as
\begin{equation}
G_{\text{eff}}(H) = \frac{G}{1 + \frac{G\tilde{c}}{\pi}H^2}.
\label{Geff_H}
\end{equation}
To simplify the notation for subsequent derivations, we define a dimensionless correction term 
\begin{equation}
\epsilon(H) \equiv \frac{G\tilde{c}}{\pi} H^2.
\end{equation}
Thus, the effective gravitational coupling can be compactly written as
\begin{equation}
G_{\text{eff}}(H) = \frac{G}{1 + \epsilon(H)}.
\label{Geff_epsilon}
\end{equation}
It is important to note that in the late-time limit where $H^2\rightarrow0$, we have $\epsilon\rightarrow0$ and $G_{\text{eff}}\rightarrow G$, ensuring the recovery of standard GR at low energy scales. In other words, the value of the gravitational constant that we measure today is very close to the $G$ that appears in the field equation of GEVAG (which is the same $G$ of GR).

Recall that the modified ``conservation law'' for energ-momentum tensor is $\nabla^a (G_{\text{eff}} T_{ab}) = 0$. The product rule of covariant derivative yields
\begin{equation}
G_{\text{eff}} \nabla^a T_{ab} + T_{ab} \nabla^a G_{\text{eff}} = 0.
\end{equation}
Let us now consider a perfect fluid described by the energy-momentum tensor $T_{ab}=(\rho + p)u_a u_b+p g_{ab}$.  where $\rho$ is the energy density, $p$ is the pressure, and $u_a$ is the four-velocity of the fluid. Projecting the conservation equation onto the comoving frame (contracting with the 4-vector $u^b$) and utilizing the normalization $u^b u_b =-1$ as usual, we obtain the modified continuity equation
\begin{equation}
\dot{\rho} + 3H(\rho + p) = -\rho \frac{\dot{G}_{\text{eff}}}{G_{\text{eff}}}.
\label{fluid_eq_general}
\end{equation}
As a consistency check, note that this is a generic result for any varying-$G$ theory, not just GEVAG. See, e.g., \cite{Hanimeli:2019wrt}.

This equation implies that the matter energy density is not independently conserved; instead, there is an energy exchange between the matter sector and the gravitational field\footnote{Strictly speaking this is also the case in GR, since a true conservation law would be $\partial^a T_{ab}=0$ instead of $\nabla^a T_{ab}=0$ \cite{Lam_2011}. The appearance of the covariant derivative already implies that the gravitational field plays a role in general. In fact, photons are red-shifted in an expanding universe, its energy is therefore not conserved independently. In GEVAG, because $G_\text{eff}$ itself is geometrical, this issue becomes more apparent. Fundamentally, the ``non-conservation'' of the matter sector is not as alarming as one may worry.}, governed by the time variation of $G_{\text{eff}}$. In the previous approaches to entropic cosmology in which $G$ remains fixed \cite{Cai:2008ys}, the RHS of the continuity equation is zero, implying the standard result $\rho\propto a^{-3(1+w)}$, where $p=w\rho$. In our model, the term on the RHS acts as a source/sink of energy, representing the energy exchange between the matter sector and the gravitational field due to the running of $G_{\text{eff}}$. This difference leads to a distinct evolution of the matter density $\rho(a)$, and consequently, a different Hubble history $H(t)$, especially in the early Universe when $\epsilon$ is large.

Using Eq.\eqref{Geff_epsilon}, we can derive the specific form of the source term for the logarithmic entropy case. Th rate of change of $G_{\text{eff}}$ is
\begin{equation}
\frac{\dot{G}_{\text{eff}}}{G_{\text{eff}}} =  - \frac{\dot{\epsilon}}{1+\epsilon}=- \frac{\epsilon}{1+\epsilon} \frac{2\dot{H}}{H},
\label{Geff_rate}
\end{equation}
since $\dot{\epsilon}=2\epsilon(\dot{H}/H)$. 
The term on the RHS represents the work done by the varying gravitational coupling. In the standard GR limit $\epsilon\rightarrow0$, the source term vanishes, and the standard continuity equation is recovered.

We now derive the equations of motion for the background geometry. The modified Einstein field equations are given by Eq.\eqref{modified_Einstein_Eq}. Assuming a flat FLRW metric $ds^2=-dt^2+a^2(t)\delta_{ab}dx^adx^b$, the $00$-component of the field equations yields the first Friedmann equation 
\begin{equation}
3H^2 = 8\pi G_{\text{eff}} \rho + \Lambda.
\label{Friedmann1}
\end{equation}
Despite the fact that this equation naively looks the same as the standard case, 
it actually has a non-trivial dependence of the Hubble parameter on the energy density. Unlike in standard cosmology, where $H^2 \propto \rho$, here $H^2$ appears on both sides of the equation (implicitly in the $G_{\text{eff}}$). This led to the structure of the dynamic equation changing from linear to nonlinear, thereby altering the physical landscape of the very early Universe.

It is important to clarify the structural difference between this framework and other modified gravity theories. In standard entropic cosmology, e.g., \cite{Cai:2008ys}, corrections are often interpreted as modifications to the spacetime geometry, effectively altering the structure of the Einstein tensor on the LHS of the field equations while assuming a constant $G$. In our GEVAG approach, we retain the standard Einstein tensor on the left-hand side. The entropy corrections are entirely encapsulated in the running gravitational coupling $G_{\text{eff}}$ on the RHS. While substituting the explicit form of $G_{\text{eff}}$ into Eq.\eqref{Friedmann1} inevitably leads to an algebraic relation involving higher powers of $H$, which is similar to geometric modifications, the physical origin is distinct. Crucially, treating the correction as a varying source coupling rather than a geometric deformation enforces a different constraint via the Bianchi identity. 

To derive the second Friedmann equation, namely the acceleration equation, we consider the trace of the field equations or directly combine the time and spatial components. Using the relation $\mathrm{~}{\ddot{a}}/{a}=\dot{H}+H^2$ and the spatial components of Eq.\eqref{modified_Einstein_Eq}, we have 
\begin{equation}
\frac{\ddot{a}}{a} = -\frac{4\pi G_{\text{eff}}}{3} (\rho + 3p) + \frac{\Lambda}{3}.
\label{Friedmann2}
\end{equation}
Again, the \emph{form} of this equation appears the same as the standard case in $\Lambda$CDM, but the devil is in the details of $G_\text{eff}$.
The system of equations \eqref{fluid_eq_general}, \eqref{Friedmann1}, and \eqref{Friedmann2} fully describes the cosmological dynamics of the logarithmic entropy corrected Universe. It can be directly verified that these equations are mathematically consistent by taking the time derivative of Eq.\eqref{Friedmann1} and using Eq.\eqref{Friedmann2} to reproduce the modified continuity equation Eq.\eqref{fluid_eq_general}.

\section{COSMOLOGICAL EVOLUTION according to GEVAG}\label{III}
In this section, we solve the modified Friedmann equations derived in the preceding section to determine the cosmological history. We analyze the evolution from the Planck era to the late-time accelerating epoch. Crucially, we interpret the dynamics through the lens of the GEVAG framework, in which deviations from standard cosmology arise due to the effective gravitational coupling $G_{\text{eff}}$ that evolves with the energy scale. Our analysis reveals that the sign of the quantum correction parameter $\tilde{c}$ leads to very different cosmological histories.

As mentioned, the first Friedmann equation Eq.\eqref{Friedmann1} establishes a non-linear relationship between the Hubble parameter $H$ and the matter energy density $\rho$. Given that $G_{\text{eff}}=G/(1+\epsilon)$ with $\epsilon = G\tilde{c}H^2/{\pi}$, we have 
\begin{equation}
H^2 = \frac{8\pi G}{3} \left[ \frac{1}{1 + \frac{G\tilde{c}}{\pi}H^2} \right] \rho + \frac{\Lambda}{3} .
\label{Master_Eq}
\end{equation}
Unlike geometric modifications, e.g., \cite{Cai:2008ys}, where extra powers of $H$ often appear on the LHS, here the modification is explicitly identified as a scaling of the source coupling. The behavior of the solution depends critically on the sign of the constant $\tilde{c}$, which dictates the sign of the correction term $\epsilon$.

\subsection{Late-Time Limit}
We first examine the late-time behavior of the Universe. In the current epoch, the Hubble parameter has the value $H\sim10^{-33}~\text{eV}$. The correction term scales as $\epsilon \sim (H/M_P)^2$. Given the immense hierarchy between the Hubble scale and the Planck scale $M_P\sim10^{28}\text{eV}$, the value of $\epsilon$ today is extremely suppressed $\epsilon_0 \sim 10^{-120}$, regardless of the coefficient $\tilde{c}$, which is generally taken to be $O(1)$.

Since $|\epsilon|\ll1$ at late time, we can analyze Eq.\eqref{Master_Eq} via a perturbative expansion. First, we re-arrange Eq.\eqref{Master_Eq} as
\begin{equation}
H^2 - \frac{\Lambda}{3} = \frac{8\pi G}{3} \rho (1+\epsilon)^{-1}.
\end{equation}
Using the binomial series expansion, we expand the term $(1+\epsilon)^{-1}$ up to the first order in $\epsilon$ as
\begin{equation}\label{e1}
H^2 - \frac{\Lambda}{3} \approx \frac{8\pi G}{3} \rho \left( 1 - \epsilon \right).
\end{equation}
Given $\epsilon_0 \sim 10^{-120}$, it is negligible compared to unity, which also means that the effective gravitational constant $G_{\text{eff}}\approx G(1-\epsilon)\rightarrow G$. Simultaneously,  the source term in the modified continuity equation Eq.\eqref{fluid_eq_general}, which depends on $\dot{G}_{\text{eff}}/G_{\text{eff}}$, effectively vanishes $\dot{\rho} + 3H(\rho+p) \approx 0$. This detailed expansion confirms that at low energies, our model naturally recovers the standard conservation law and the standard $\Lambda$CDM cosmology. The tiny value of $\epsilon$ ensures that the modifications to gravity do not affect the late-time Universe, and thus would have no issues with various observational constraints like solar system tests. 

It is tempting to look at Eq.\eqref{e1} and hope that even with $\Lambda=0$ one might obtain a dark energy from the correction term by setting $\Lambda_\text{eff}\equiv -8\pi G \rho \epsilon$. However, this is not valid because we know that $\rho$ in the matter sector has the same order of magnitude as $\rho_\Lambda$. Such a $\Lambda_\text{eff}$ would be far too small. 

\subsection{Very-Early Universe}
The distinct features of the GEVAG model emerge in the early Universe where energies approach the UV and $\epsilon$ becomes significant. In this regime, $\rho \gg \Lambda$, so we shall neglect the cosmological constant. Eq.\eqref{Master_Eq} simplifies to $3H^2 = 8\pi G_{\text{eff}}  \rho$. The dynamics are entirely controlled by the behavior of the denominator in $G_{\text{eff}}$. For the early stage of the Universe, we distinguish between two scenarios. 

\subsubsection{$\tilde{c}<0$: Density Saturation}
The first case is $\tilde{c}<0$, a scenario that is favored by LQG. Let us explicitly write $\tilde{c}=-|\tilde{c}|$, and the effective gravitational constant becomes $G_{\text{eff}} = \frac{G}{1 - \frac{G|\tilde{c}|}{\pi}H^2}$.
As the radius contracts and $H$ increases,  the term $\frac{G|\tilde{c}|}{\pi}H^2$ grows. This implies a fundamental bound on the matter energy density, which prevents the Universe from collapsing into a singularity.

To determine the maximum density, we re-arrange Eq.\eqref{Master_Eq} into a form suitable for finding the critical point (the extremum of the function $\rho$):
\begin{equation}
\frac{8\pi G}{3} \rho = H^2 \left( 1 - \frac{G|\tilde{c}|}{\pi}H^2 \right).
\label{bounce_potential}
\end{equation}
Differentiating this function, one can easily find the critical Hubble parameter squared as $H^2_c = \frac{\pi}{2G|\tilde{c}|}$. At this point, the correction term is exactly $\epsilon(H_c)=-\frac{G|\tilde{c}|}{\pi}\left(\frac{\pi}{2G|\tilde{c}|}\right)=-1/2$, which in turn implies that the maximum effective gravitational coupling is $G_{\text{eff}}(H)=2G$. Substituting these back into Eq.\eqref{bounce_potential} to find the critical density gives
\begin{equation}
\rho_{\text{crit}} = \frac{3}{8\pi G} \left( \frac{\pi}{4G|\tilde{c}|} \right) = \frac{3}{32 G^2 |\tilde{c}|}.
\label{rho_crit}
\end{equation}
This result implies that the density is strictly bounded by $\rho_{\text{crit}}$. The classical Big Bang singularity $\rho_{\text{crit}}\to \infty$ is dynamically forbidden because the modified Friedmann constraint imposes a density saturation. It provides a thermodynamic realization of the discreteness of spacetime predicted by LQG, acting as a kinematic cutoff in phase space.

Note that Ref.\cite{Cai:2008ys} by Cai et al. derived a modified Friedmann equation in their Eq.(16), but the implications for the maximum density were not explicitly explored. A re-examination of their equation reveals a density bound\footnote{Here we retain their notation for reference. In fact, their $\alpha$ is equivalent to our $\tilde{c}$. }  $\rho_{\mathrm{max}}^{\mathrm{Cai}}=\frac{3}{16G^2|\alpha|}$ with a critical Hubble parameter squared being $H_c^2 = -\pi/(\alpha G)$, which differs from our GEVAG result by exactly a factor of 2. This discrepancy arises from the different mathematical procedures employed to derive the Friedmann equation. In Ref.\cite{Cai:2008ys}, the equation is derived by integrating the differential form of the Clausius relation $\int H^2~\text{d}H^2$, which inherently introduces a factor of $1/2$ in the higher-order geometric term. In contrast, our GEVAG framework treats the entropy correction as a direct algebraic rescaling of the effective gravitational coupling $G_{\text{eff}}$ within the field equations, which generates the quartic term without the need for an integration factor.




An intriguing yet problematic feature of the constant-$G$ entropic correction approach emerges when examining the dynamics near the Planck scale. Refer to the modified acceleration equation $\dot{H} \left(1 + \frac{\alpha G}{\pi} H^2\right) = -4\pi G (\rho + p)$ in Ref.\cite{Cai:2008ys}, the factor $ \left(1 + \frac{\alpha G}{\pi} H^2\right) $ multiplying $\dot{H}$ can vanish for $ \alpha < 0 $ at $ H_c^2 = -\pi/\alpha G$. This induces a divergence in the Hubble acceleration $ \dot{H} $ at a finite energy density, namely the so-called ``sudden singularity'' \cite{Barrow:2004xh, Nojiri:2005sx}, which is generally taken to be physically untenable and indicates a breakdown of the thermodynamic description in this regime. Sudden singularities are known to occur in the context of loop quantum cosmology (LQC). In fact, the authors of Ref.\cite{Cailleteau:2008wu} argued that the effective equations of LQC are not in themselves sufficient to avoid the occurrence of singularities. 
The GEVAG approach does not have this shortcoming; the acceleration equation Eq.\eqref{Friedmann2} is obtained directly from the spatial component of the field equation as $\dot{H}+H^2\sim G_{\mathrm{eff}}(\rho+3p)$, having no divergent $ \dot{H} $ for a finite $G_{\mathrm{eff}}$.


\subsubsection{$\tilde{c}>0$: Asymptotic Safety}
We now consider the case where $\tilde{c}>0$, which is often discussed in the context of ASG. In this scenario, $\epsilon>0$. For very large densities at early time, the quartic term in the expansion of Eq.\eqref{Master_Eq} dominates
\begin{equation}
\frac{8\pi G}{3} \rho = H^2 + \frac{G\tilde{c}}{\pi} H^4 \approx \frac{G\tilde{c}}{\pi} H^4.
\end{equation}
Solving for $H$, we find the high-energy scaling relation
\begin{equation}
H \approx \left( \frac{8\pi^2}{3\tilde{c}} \right)^{1/4} \rho^{1/4}.
\end{equation}
This scaling $H \propto \rho^{1/4}$ contrasts sharply with the classical GR scaling $H \propto \rho^{1/2}$. In this limit, the effective gravitational constant behaves as
\begin{equation}
G_{\text{eff}} \approx \frac{G}{\epsilon} = \frac{\pi}{\tilde{c} H^2} \rightarrow 0.\label{UVasg}
\end{equation}
The vanishing of $G_{\text{eff}}$ in the UV limit realizes the asymptotic safety scenario, where the theory flows to an NGFP with a finite dimensionless coupling $g_* \equiv G_{\text{eff}} H^2 \rightarrow \pi/\tilde{c}$. 

The fact that $H \propto \rho^{1/4}$ is consistent with ASG can also be seen as follows: In ASG, we have $G(k)=g_*/k^2$ where $g_*$ is a fixed-point constant and $k$ is some energy scale. In cosmology, a commonly used identification is $k \sim H$ as was done in the preceding paragraph; see also \cite{Ahn:2011qt}. This implies $G \sim H^{-2}$. The usual Friedmann equation thus leads to
\begin{equation}
H^2 = \frac{8\pi G(k)}{3}\rho = \frac{8\pi}{3H^2}\rho,
\end{equation}
or equivalently $H^4 \sim \rho$.


Note that in ASG, at the exact fixed point, scale invariance is restored, and the Universe typically approaches a quasi–de Sitter geometry with a constant Hubble parameter $H_*$. In this limit, the Ricci scalar $R = 6(2H^2 + \dot{H})= 12H_*^2$ remains finite. The fixed point thus acts as a UV attractor that prevents the trajectory from flowing into a curvature singularity, suggesting that curvature saturates at a finite value rather than diverging as in classical theory.

There are, however, some concerns in this scenario of $\tilde{c}>0$. Notably, a naive extrapolation of the entropy formula $S=A/4G+\tilde{c}\ln(A/G)$ to the limit $H\to\infty\mathrm{~(}A\to0\mathrm{)}$ leads to a divergent and negative entropy: $S\to-\infty$. However, this mathematical limit is physically inaccessible because the logarithmic correction contains an intrinsic scale\footnote{Here it should be argued that the logarithmic correction term is only the leading-order correction. Higher-order terms may be important when $A$ is small. See paragraph around Eq.(29) of \cite{Ong:2025ent}.}, namely the Planck area $A_p=G$. It is physically compelling to postulate that the effective thermodynamic description is valid only down to this natural UV cutoff, i.e.,
$A\gtrsim G$. At this cutoff scale, the logarithmic term vanishes, and the entropy is of order unity
\begin{equation}
S_{\text{initial}} \approx \frac{G}{4G} + \tilde{c} \ln(1) \sim \mathcal{O}(1).
\end{equation}
This interpretation resolves the apparent divergence and yields a profound cosmological picture, i.e., the ASG branch also originates from a low-entropy state, which is consistent with the conclusion of Ref.\cite{Bonanno:2007wg}. This reminds us of the ``arrow of time" issue, and one can refer to Refs.\cite{Greene:2009tt, Sloan:2018osd} for some attempts to resolve the arrow of time problem by weakening gravity in early time.


\section{Discussion on Inflationary Conditions}\label{inflation}
The preceding analysis reveals that the sign of the coefficient $\tilde{c}$ determines the asymptotic strength of gravity in the early Universe, leading to two distinct cosmological histories. In this section, we shall continue with investigating how this gravitational strength directly governs the naturalness of initiating a subsequent period of slow-roll inflation. The crucial point lies in how the effective gravitational coupling $G_{\text{eff}}$ modifies the slow-roll conditions for an inflaton field.

Consider the simplest standard inflationary scenario driven by a homogeneous scalar field $\phi$ with a potential $V(\phi)$. The energy density and pressure are given, respectively, by
 \begin{equation}
\rho_\phi = \frac{1}{2} \dot{\phi}^2 + V(\phi), \quad p_\phi = \frac{1}{2} \dot{\phi}^2 - V(\phi).
\end{equation}
In the slow-roll regime, the potential energy dominates, $\dot{\phi}^2 \ll V(\phi)$, leading to $\rho_\phi \approx V$. 

It is crucial to emphasize that the equation of motion that governs the inflaton field is now changed by the modified continuity equation Eq.\eqref{fluid_eq_general}. Substituting the energy density and pressure into Eq.\eqref{fluid_eq_general}, one can get the modified equation  of motion as
\begin{equation}
\dot{\phi}\ddot{\phi}+V^{\prime}\dot{\phi}+3H\dot{\phi}^2=-\frac{\dot{G}_{\mathrm{eff}}}{G_{\mathrm{eff}}}\left(\frac{1}{2}\dot{\phi}^2+V\right).
\label{mkg}
\end{equation}
Applying the slow-roll approximation, we can neglect the acceleration term $\ddot{\phi}$ and $\dot{\phi}^2 \ll V(\phi)$, we obtain 
\begin{equation}
V^{\prime}\dot{\phi}+3H\dot{\phi}^2\approx-\rho\frac{\dot{G}_{\mathrm{eff}}}{G_{\mathrm{eff}}}. \label{mkg2}
\end{equation}
In the UV limit of the $\tilde{c}>0$ branch, the effective gravitational constant behaves as Eq.\eqref{UVasg}, thus leading to
\begin{equation}
\frac{\dot{G}_{\mathrm{eff}}}{G_{\mathrm{eff}}}\approx-2\frac{\dot{H}}{H},
\end{equation}
and given the slow-roll requirement that $\rho_\phi \approx V$, Eq.\eqref{mkg2} can be expressed as
\begin{equation}
V^{\prime}\dot{\phi}+3H\dot{\phi}^2\approx2V\frac{\dot{H}}{H}.
\label{mkg3}
\end{equation}

To demonstrate the naturalness of initiating inflation, we introduce the first slow-roll parameter\footnote{The second slow-roll parameter $\eta_{H}$ is crucial for determining the duration and detailed observational predictions of inflation. We shall only focus on the first slow-roll parameter to investigate the conditions for the Universe to enter an inflationary phase.} $\epsilon_H \equiv -\dot{H}/H^2$, which is precisely the measure that governs this onset. (Do not confuse $\epsilon_{H}$ with $\epsilon$, the latter being the correction term in $G_\text{eff}$.)

The modified Friedmann equation \eqref{Friedmann1} with $\Lambda=0$ is
\begin{eqnarray}
H^2 + \frac{G\tilde{c}}{\pi} H^4 = \frac{8\pi G}{3}V.\label{f3}
\end{eqnarray}
In the ASG branch ($\tilde{c}>0$), the $H^4$ term dominates Eq.\eqref{f3} in the UV limit, yielding
\begin{equation}
H^4 \approx \frac{8\pi^2}{3\tilde{c}} V.
\label{asga}
\end{equation}
As previously discussed, this scaling behavior also follows from the NGFP regime of asymptotically safe gravity, where $G(k)\sim g_*/k^2$ and $k\sim H$, leading to $H^4\propto V$ in the RG-improved Friedmann equation. (In addition to $k \sim H$, there are other possible choices of $k$, see the discussions in \cite{Bonanno:2001xi}, in which the authors chose to work with $k \sim 1/t$. For power-law expansion, this is the same as $k \sim H$.)

We can get a relation $\dot{H}/H\approx V^{\prime}\dot{\phi}/4V$ by using Eq.\eqref{asga}. Substituting $\dot{H}/H$ back to the RHS of Eq.\eqref{mkg3}, we next obtain
\begin{equation}
V^{\prime}\dot{\phi}+3H\dot{\phi}^2\approx2V\left(\frac{V^{\prime}\dot{\phi}}{4V}\right)=\frac{1}{2}V^{\prime}\dot{\phi},
\end{equation}
from which we deduce that
\begin{equation}
\dot{\phi}\approx-\frac{V^{\prime}}{6H}.
\end{equation}
This should be compared to the standard result in GR, which is $-\frac{V^{\prime}}{3H}$. Here, the denominator has changed to $6H$, which means that at the UV limit, the friction term doubles. This is the dynamic effect brought about by the varying-$G$.

Using the accelerating equation $\dot{H}=-4\pi {G}_{\mathrm{eff}}\dot{\phi}^2$ for the inflaton field with Eq.\eqref{asga}, we can obtain the expression for the first slow-roll parameter in the deep UV regime
\begin{equation}
\epsilon_H = -\frac{\dot{H}}{H^2} \approx \left( \frac{1}{48\pi} \sqrt{\frac{3\tilde{c}}{2}} \right) \frac{(V')^2}{V^{3/2}}. \label{slow1}
\end{equation}
This result differs from the standard GR expression, where $\epsilon_H^{\text{GR}} \approx \frac{1}{16\pi G} \left( \frac{V'}{V} \right)^2$. Recall the condition $\epsilon_H \ll 1$. Given $\rho_\phi \approx V$ is a sub-Planckian energy scale which is less than $1$, this condition requires that $V'$ must be smaller, that is, the potential energy $V(\phi)$ must be extremely flat. However, the dependence on $V^{3/2}$ in the denominator of the ASG expression means that the potential $V$ naturally relaxes the constraint on the slope $V'$, allowing for a steeper potential with less fine-tuning. This makes the onset of slow-roll inflation a more natural outcome in the ASG-like branch (namely, the $\tilde{c}>0$ branch). We emphasize that from the point of view of ASG, this result is obtained after mapping the RG-improved cosmological evolution onto an effective scalar-field description. The scalar field introduced in this way should be viewed as a convenient parametrization of the RG-driven dynamics rather than a fundamental degree of freedom. In the standard asymptotic safety scenario, inflation can emerge purely from the scale dependence of the gravitational couplings without requiring a fundamental inflaton field \cite{Bonanno:2001xi, Bonanno:2017pkg}.

The fact that inflation is less fine-tuned in the $\tilde{c}>0$ case is consistent with our preceding finding that this branch has a weaker effective gravitational constant in the very early Universe, which ameliorates the arrow of time problem. The role of inflation in the context of the arrow of time problem has long been debated. For example, it has been argued that inflation alone cannot resolve the problem \cite{Page:1983uh} because it also requires fine-tuned initial conditions to start, and thus one is only pushing low-entropy initial conditions to earlier times. Given that the $\tilde{c}>0$ branch is expected to have more natural initial conditions, it is no surprise that it also more readily allows slow-roll inflation to occur.   

Conversely, in the $\tilde{c}<0$ case favored by LQG, the situation is fundamentally different. 
Recall the slow-roll approximation \eqref{mkg2}, we can check that the RHS of the source term is a second-order infinitesimal, because $\dot{G}_{\text{eff}}/G_{\text{eff}}\propto \dot{H}$ from Eq.\eqref{Geff_rate} with the accelerating relation $\dot{H}=4\pi G_{\text{eff}} (\rho+p)  \propto \dot{\phi}^2$. So we can neglect the source term under the lowest-order approximation, and we get a GR-like relation $\dot{\phi}\approx-{V^{\prime}}/{3H}$.
Therefore the first slow-roll parameter for the $\tilde{c}<0$ branch is 
\begin{equation}
\epsilon_H = -\frac{\dot{H}}{H^2} \approx \frac{4\pi {G}_{\text{eff}}}{9} \frac{(V')^2}{H^4},
\end{equation}
and one can find that it is consistent with the form of GR where $\epsilon_H^{\text{GR}} \approx \frac{4\pi G}{9} \frac{V'^2}{H^4} $. When approaching the critical point, ${G}_{\text{eff}} \to 2G$, which indicates that the potential energy $V(\phi)$ might be flatter than the GR case, but in general, these two cases yield the same order of magnitude for $\epsilon_H$.

Based on the above discussions, we can conclude that logarithmic corrections with different signs will lead to very different behaviors of the effective gravitational constant ${G}_{\text{eff}}$. In particular, the modified cosmological dynamics suggest different initial conditions for inflation, although it cannot be conclusively determined which scenario is better (though $\tilde{c}>0$ has the advantage from the point of view of the arrow of time problem, it does not mean that theories with $\tilde{c}<0$ cannot resolve the arrow of time issues in other ways), it offers a new perspective for entropic cosmology.

\section{THERMODYNAMIC CONSISTENCY AND Parameter range}\label{IV}
\subsection{Generalized Second Law}
Considering the Jacobson method is established quasi-locally, we shall verify the thermodynamic consistency of the model by checking the generalized second law (GSL): $\dot{S}_{\mathrm{total}}=\dot{S}_h+\dot{S}_m \geq 0$, where $S_h$ and $S_m$ are the horizon and contained matter entropy, respectively.

First, we calculate the rate of change of the horizon entropy
\begin{equation}
\dot{S}_h = \frac{\text{d}S_h}{\text{d}A} \dot{A} =  - \frac{2\pi \dot{H}}{G_{\text{eff}} H^3}.
\end{equation}
Assuming a quasi-local horizon thermodynamic description\footnote{Here we must be very cautious. We adopt the quasi-local horizon thermodynamic description, not assuming that the cosmic fluid is in strict thermal equilibrium. But recently, from the aspect of open quantum systems, it has been shown that the de Sitter period satisfies the Kubo-Martin-Schwinger (KMS) condition, a hallmark of thermal equilibrium \cite{Alicki:2023rfv}, and some novel phenomena will happen following this hypothesis \cite{Alicki:2023tfz, Wu:2025usd}, which effectively allows the use of a horizon temperature. Consequently, this provides a potentially consistent framework for both late-time cosmology and inflationary epochs.} \cite{Cai:2005ra, Cai:2006rs}, the apparent horizon temperature is $T=H/2\pi$, and the heat flux is
\begin{equation}
T \dot{S}_h = \left( \frac{H}{2\pi} \right) \left( - \frac{2\pi \dot{H}}{G_{\text{eff}} H^3} \right) = - \frac{\dot{H}}{G_{\text{eff}} H^2}.
\end{equation}
For the matter sector, we need to use the modified continuity equation Eq.\eqref{fluid_eq_general} derived in Sec.(\ref{II}). Substituting said continuity equation into the Gibbs equation\footnote{Here the Gibbs relation is employed in an effective, coarse-grained sense to encode the energy exchange of the cosmic fluid, not a true microscopic thermal equilibrium.} $T\dot{S}_m=V\dot{\rho}+(\rho+p)\dot{V}$ where $V=\frac{4}{3}\pi R_H^3$, we get
\begin{equation}
T \dot{S}_m = V \left[ -3H(\rho+p) - \rho \frac{\dot{G}_{\text{eff}}}{G_{\text{eff}}} \right] - 3HV(\rho+p).
\end{equation}
Then using the Friedmann acceleration relation $\dot{H}=-4\pi G_\mathrm{eff}(\rho+p)$, we can obtain
\begin{equation}
T \dot{S}_m = \frac{\dot{H}}{G_{\text{eff}} H^2} + \frac{\dot{H}^2}{G_{\text{eff}} H^4} - V\rho \frac{\dot{G}_{\text{eff}}}{G_{\text{eff}}}.
\end{equation}
Summing the horizon and matter contributions, we find that the leading geometric terms $\dot{H}/(G_{\text{eff}} H^2)$ cancel out exactly, which results in
\begin{equation}
T \dot{S}_{\text{total}} = \frac{\dot{H}^2}{G_{\text{eff}} H^4} - V\rho \frac{\dot{G}_{\text{eff}}}{G_{\text{eff}}}.
\end{equation}
Using the relation $V\rho=\frac{4}{3}\pi R_H^3(\frac{3 H^2}{8 \pi G_{\text{eff}}})=\frac{1}{2HG_{\mathrm{eff}}}$ and substituting the expression Eq.\eqref{Geff_rate}, we arrive at the condition
\begin{equation}
T \dot{S}_{\text{total}} = \frac{\dot{H}}{G_{\text{eff}} H^2} \left[ \frac{\dot{H}}{H^2} + \frac{\epsilon}{1+\epsilon} \right].
\label{GSL_condition}
\end{equation}

Firstly, recall our estimate in Sec.(\ref{III}) that $\epsilon_0 \sim 10^{-120}$ at late time, so the late-time Universe is essentially just $\Lambda$CDM, for which the GSL is known to be valid. Since $\epsilon_0$ is so tiny, the sign of the coefficient $\tilde{c}$ does not matter. Even during the period of matter/radiation-domination, this conclusion remains true. To see this, note that the acceleration relation from Eq.\eqref{Friedmann2} gives $\dot{H}=-4\pi G_\mathrm{eff}\rho(1+w)= -\frac{3}{2}H^2(1+w)$, leading to $\frac{\dot H}{H^2}= -\frac{3}{2}(1+w)$. We have $\dot H/H^2\sim-1.5$ at the matter-dominated period and $\dot H/H^2\sim-2$ at the radiation-dominated period. Considering $H\ll M_p$ during the above periods, the order of magnitude of the correction term $\epsilon/(1+\epsilon)$ in Eq.(\ref{GSL_condition}) is completely negligible compared to that of the leading term $\dot H/H^2$, and thus GSL is always satisfied.


The situation in the early Universe is subtle. For the $\tilde{c}<0$ branch, the sign of $\epsilon/(1+\epsilon)$ is always negative given that $\epsilon<0$ and has a floor value at density saturation with $\epsilon(H_c)=-1/2$. Considering a typical slow-roll inflation, the Hubble parameter variation rate is negative, i.e., $\dot {H} <0$, which ensures that $T \dot{S}_{\text{total}}>0$ always holds, thereby obeying the GSL. For the $\tilde{c}>0$ branch, the GSL does not hold automatically during inflation but imposes specific constraints on the dynamics of inflation. In the low-energy scale regime $\epsilon\ll1$, and $\epsilon/(1+\epsilon)\sim\epsilon$. From Eq.(\ref{GSL_condition}), we can see that the validity of GSL now requires $\epsilon_H \gtrsim \epsilon$. For a sub-Planckian potential energy, this condition is likely to be satisfied, unless the potential energy is extremely flat. When close to the UV fixed point, $\epsilon/(1+\epsilon)\sim\mathcal{O}(1)$, the slow-roll condition must be violated such that $\epsilon_H\sim\mathcal{O}(1)$ for GSL to be satisfied, which favors a lower energy scale. 

We must emphasize that these discussions on the early Universe are based on an effective quasi-local horizon thermodynamic description. In fact, the early Universe might involve complex non-equilibrium processes that went far beyond such a description. Thus even if GSL is violated at the UV fixed point for ASG-like scenarios, it is unclear if this truly signals a pathology.

\subsection{Possible Parameter Range}
In the previous part of the article, following the conventions of the literature on ASG and LQG as well as other quantum gravity approaches, we consistently treated the parameter $\tilde{c}$ as an order-unity coefficient. In this part, we will discuss if observations can give some constraints to its order of magnitude.

Firstly, we use observational constraints from Big Bang nucleosynthesis (BBN), which occurs at a temperature of $T\sim 1 \, \text{MeV}$. The success of the standard BBN predictions for light element abundances imposes strict limits on any deviation from the gravitational constant $G$. In the radiation-dominanated era, $H \simeq 1.66\,\sqrt{\bar{g}_*}\,(T^2/M_{\rm Pl})$, and $\bar{g}_*\simeq 10.75$ is the effective degrees of freedom\footnote{This is usually denoted as $g_*$ in cosmological literature, but we have used $g_*$ in the ASG context, hence a slight change of notation to avoid confusion.} for typical BBN \cite{Husdal:2016haj}, leading to $H_{\rm BBN}^2\simeq 2.0\times10^{-49}\,{\rm GeV}^2$ during BBN. Considering $G=M_{\rm Pl}^{-2}\simeq 6.7\times10^{-39}\,{\rm GeV}^{-2}$, we have $\epsilon_{\rm BBN}\simeq\tilde{c}\cdot 4.3\times10^{-88}$. 

Ref.\cite{Alvey:2019ctk} suggested $G_{\mathrm{BBN}}/G_{0}=0.98_{-0.06}^{+0.06}$ and indicated a $2\%$ upper limit of deviation in G. Then, for our model \eqref{Geff_H}, $(G_{\rm eff}-G)/G= -\epsilon/(1+\epsilon)$. Consequently, this suggests an extremely loose upper bound $|\tilde c|\lesssim 1\times10^{86}$ (this bound is even looser than the best bound of GUP of gravitational origin ($\lesssim 10^{60}$) \cite{Feng:2016tyt, Scardigli:2019pme}). This also means that the BBN constraint is not sensitive unless $\tilde{c}$ is extremely large. It is worth noting that the terrestrial local measurements report that $\delta G/G$ less than $\sim\mathcal{O}(10^{-8})$ at late time \cite{Dai:2021jnl}.

While late-time observations allow for a wide range of $\tilde{c}$, the physics of the Planck era suggest a small range of the parameter selection. For the $\tilde{c}<0$ branch favored by LQG, the critical density is given by $\rho_{\text{crit}} = 3/(32 G^2 |\tilde{c}|)$. Demanding that this density matches the prediction from LQC $\rho_{\text{crit}}^{\text{LQC}} \approx 0.41 \rho_p$ \cite{Ashtekar:2007em}, where $\rho_p$ is the Planck density, we have $|\tilde{c}| \approx 0.23$. This value is consistent with the values of $\tilde{c}$ derived from black hole entropy counting in LQG ($\tilde{c} = -1/2$ or other $O(1)$ values). This quantitative agreement supports the interpretation of our model as an effective thermodynamic description of LQC.

For the  $\tilde{c}>0$ branch favored by ASG, we argued in Sec.(\ref{III}) that the validity of the theory requires a natural UV cutoff at the Planck area $A_p\sim G$. At this scale, the logarithmic term vanishes, ensuring a finite, low-entropy initial state. This physical condition implies that the correction term $\epsilon \sim \tilde{c} (H/M_p)^2$ becomes of order unity exactly at the Planck scale $H \sim M_p$, leading to $\tilde{c} \sim \mathcal{O}(1)$. This provides a theoretical upper bound. If $\tilde{c}$ were excessively large, e.g., $10^{10}$, the cutoff would occur at energies far below the Planck scale, conflicting with standard particle physics. If $\tilde{c}$ were extremely small, we would have the undesired situation whereby the asymptotic safety behavior would not activate until trans-Planckian energies. Thus, $\tilde{c} \sim \mathcal{O}(1)$ is a natural prediction for the asymptotic safety scenario as well.

\section{Conclusion}\label{V}

In this work, we investigated the different cosmological behaviors when the quantum-gravitational logarithmic correction was incorporated into the GEVAG framework. Depending on the sign of the correction term (the coefficient $\tilde{c}$), the behaviors are quite different. GEVAG takes into account how any modification to the area law $S=A/4G$ affects gravity, the result of which is an effective gravitational ``constant'' $G_\text{eff}$ that is a function of the apparent horizon area. For the $\tilde{c}<0$ case favored by LQG, $G_\text{eff} \to 2G$ at very early times, and the inflationary condition is similar to the GR case. The GEVAG approach does still have an advantage over the standard fixed $G$ assumption, namely it could avoid a sudden singularity from happening. In the $\tilde{c}>0$ favored by ASG, we argued that $G_\text{eff} \to 0$ in early time, which could ameliorate the arrow of time problem. In fact, slow-roll inflation seems to be more natural in this case. 

It should be mentioned that the sign of $\tilde{c}$ is also of great importance in the context of the generalized uncertainty principle (GUP). For example, $\tilde{c}<0$ corresponds to a positive GUP parameter (see \cite{Ong:2025ent}). The $\tilde{c}>0$ case corresponds to a negative GUP parameter, a possibility that was previously argued for based on the existence of the Chandrasekhar limit of white dwarfs (a positive GUP parameter seemingly removes the Chandrasekhar limit) \cite{Ong:2018zqn}. 

The GEVAG framework is still new and thus has many potential applications that could be investigated in the future. One possibility is to look into how GEVAG influences gravitational collapse, perhaps along the lines of Ref.\cite{Hassannejad:2024cbu}, which has also investigated the case of ASG. We hope that future research may reveal more surprising results brought by GEVAG.

%

\end{document}